\pdfoutput=1
\documentclass[prd,preprint,floats,floatfix,aps,nofootinbib,preprintnumbers]{revtex4}
\usepackage{graphics,graphicx,psfrag}
\usepackage{amsmath,amssymb}
\usepackage[sort&compress]{natbib}
\usepackage{bm}	% for boldface in math
\usepackage{verbatim}
\usepackage{multirow}
\usepackage{subfigure}
\usepackage{ulem}
\usepackage{color}
\usepackage{slashed}
\usepackage[linktocpage=true,bookmarksnumbered=true,colorlinks=true,plainpages,linkcolor=blue,citecolor=red]{hyperref}
\usepackage[utf8]{inputenc}  %per INSPIRE directions

\graphicspath{{./figs/}}

\newcommand{\invfb}{\text{fb}^{-1}}

\begin{document}
\preprint{
{\vbox {
\hbox{\bf MSUHEP-17-011}
\hbox{\today}
}}}
\vspace*{2cm}

\title{Broadening the Reach of \\ Simplified Limits on Resonances at the LHC}
\vspace*{0.25in}   % makes space between address and abstract
\author{R. Sekhar Chivukula$^1$}
\email{sekhar@msu.edu}
\author{Pawin Ittisamai$^2$}
\email{ittisama@msu.edu}
\author{Kirtimaan Mohan$^1$}
\email{kamohan@pa.msu.edu}
\author{Elizabeth H. Simmons$^1$}
\email{esimmons@msu.edu}
\affiliation{\vspace*{0.1in}
$^1$ Department of Physics and Astronomy\\
Michigan State University, East Lansing U.S.A.\\
$^2$ Department of Physics, Faculty of Science\\
Chulalongkorn University, Bangkok 10330, Thailand
}
%\vspace*{0.25 in} % makes space between address and abstract

\begin{abstract}
\vspace{0.5cm}
\noindent
Recently, we introduced an approach for more easily
interpreting searches for  resonances at the LHC -- and to aid in distinguishing between realistic
and unrealistic alternatives for potential signals. This ``simplified limits"
approach was derived using the narrow width approximation (NWA) -- and therefore was not
obviously relevant in the case of wider resonances. Here, we broaden the
scope of the analysis. First, we explicitly generalize the formalism to encompass
resonances of finite width. We then examine how the width of the resonance modifies
bounds on new resonances that are extracted from LHC searches. Second, we demonstrate, using a wide variety of cases,
with different incoming partons, resonance properties, and decay signatures, that the 
limits derived in the NWA yield pertinant, and somewhat
conservative (less stringent) bounds on the model parameters. We conclude that the original simplified limits
approach is useful in the early stages of evaluating and
interpreting new collider data and that the generalized approach is a valuable further aid when evidence points toward a broader resonance.
\end{abstract}

\maketitle

%%%%%%%%%%%%%%%%%%%%%%%%%%%%%%%%
%  INTRODUCTION
%%%%%%%%%%%%%%%%%%%%%%%%%%%%%%%%

\section{Introduction}

New physics searches at the LHC commonly explore two-body scattering processes for signs that a resonance arising from physics Beyond the Standard Model (BSM) is being produced in the $s$-channel and immediately decaying to visible final state particles.  Observed limits on the production cross-section ($\sigma$) times branching fraction ($BR$) for the process as a function of the resonance mass are compared with the predictions of a few benchmark models, each corresponding to one choice of spin, electric charge, weak charge, and color charge for the new resonance, evaluated for specific parameter values.  However, for a given choice of spin and charges there will actually be multiple detailed theoretical realizations corresponding to  different strengths and chiralities of the resonance's couplings to initial-state partons and to decay products.  The benchmarks shown in the analyses often correspond to convenient examples that have large production rates (like a leptophobic $Z'$ boson) or are already encoded in available analysis tools.  

In recent work~\cite{Chivukula:2016hvp}, we argued that when first evaluating new results, especially if signs of a small excess exist, it would be valuable to compare the data with entire classes of models, to see whether any resonances with particular production modes and/or decay patterns (e.g., a spin-zero state produced through gluon fusion and decaying to diphotons) could conceivably be responsible for a given deviation in cross-section data relative to standard model predictions. Using a simplifed model of the resonance allowed us to convert an estimated signal cross section into bounds on the product of the branching ratios corresponding to production and decay. This quickly reveals whether a given class of models could possibly produce a signal of the required size at the LHC and circumvents the present need to make laborious comparisons of many individual theories with the data one by one.  Moreover, the ``simplified limits variable" $\zeta$, which factors in the width-to-mass ratio of the resonance, produces even more compact and easily interpretable results.

We began by establishing a general framework for obtaining simplified limits and outlining how it applies for narrow resonances with different numbers of production and decay modes.  We then analyzed applications of current experimental interest, including resonances decaying to dibosons, diphotons, dileptons, or dijets.  We further illustrated how easy it was to compare the calculated value of the simplified limits variable $\zeta$ for a specific instance of a new state with the experimental upper bound on $\zeta$ in order to determine whether that particular instance was a viable candidate to explain the excess. 

Here, we report on how to broaden the ``simplified limits" approach of \cite{Chivukula:2016hvp}. After all, new physics may appear as a scattering excess that is not obviously due to a narrow s-channel resonance.  We are therefore generalizing our simplified limits framework to handle resonances of moderate width treated in the Breit-Wigner approximation.  

Our generalized method addresses the implications of any signs of a small excess, as well as indicating how to interpret experimental exclusion curves.  It builds upon our previous results for identifying the color \cite{Atre:2013mja,Chivukula:2014npa} and spin \cite{Chivukula:2014pma} properties of new resonances decaying to dijet final states, extending them to a wider variety of final states and to situations in which only a small deviation possibly indicative of a resonance has been observed.  This contrasts with studies in the literature that have focused on the discovery reaches for multiple $Z'$ models at a single collider  \cite{Carena:2004xs}, compared discovery reaches across multiple colliders \cite{Dobrescu:2013coa}, or assessed the potential reach of proposed new colliders \cite{Eichten:1984eu}.  Recent work \cite{Franceschini:2015kwy} more similar in spirit to ours has focused specifically on a potential 750 GeV diphoton signal at the LHC \cite{ATLAS-Diphoton,Moriond-ATLAS,ATLAS-CONF-2016-018,CMS:2015cwa,CMS:2015dxe,Moriond-CMS,CMS-PAS-EXO-16-018}. 

 In the next section, we will briefly review the key results from our work on narrow resonances.  Section III discusses how we extend these to broader resonances, treated in the Breit-Wigner approximation.  Section IV discusses applications to broader resonances decaying to dileptons, dibosons, and dijets.  The final section presents our conclusions.

%%%%%%%%%%%%%%%%%%%%%%%%%%%%%%%%
%  SIMPLIFIED LIMITS FOR NARROW RESONANCES
%%%%%%%%%%%%%%%%%%%%%%%%%%%%%%%%

\section{Recap: Simplified Limits on Narrow Resonances}

In \cite{Chivukula:2016hvp,Chivukula:2017qsi} we proposed a general method for quickly determining whether a small excess observed in collider data could potentially be attributable to the production and decay of a single, relatively narrow, s-channel resonance belonging to a generic category, such as a leptophobic $Z^\prime$ boson or a fermiophobic $W^\prime$ boson. Using a simplifed model of the resonance allows us to convert an estimated signal cross section into bounds on the product of the branching ratios corresponding to production and decay.  Moreover, the ``simplified limits variable" $\zeta$, which factors in the width-to-mass ratio of the resonance, produces even more compact and easily interpretable analyses. Here we mention a few key results that set the context for our present work.

The tree-level partonic production cross-section for an arbitrary 
$s$-channel resonance $R$ produced by collisions of particular initial state partons $i,j$ and decaying to  a single final state $x,y$ at the LHC can be written \cite{Harris:2011bh,Agashe:2014kda}
\begin{equation}
\hat{\sigma}_{ij\to R\to xy}(\hat{s}) = 16 \pi (1 + \delta_{ij}) \cdot {\cal N} \cdot
\frac{\Gamma(R\to i+j) \cdot \Gamma(R\to x+y)}
{(\hat{s}-m^2_R)^2 + m^2_R \Gamma^2_R} ~ \qquad {\cal N} = \frac{N_{S_R}}{N_{S_i} N_{S_j}} \cdot
\frac{C_R}{C_i C_j},
\label{eq:nr}
\end{equation}
where $N_S$ and $C$ count\footnote{While ${\cal N}$ depends on the color and spin properties of the incoming partons $i,j$, in most cases \cite{Chivukula:2016hvp} this factor is the same for all relevant production modes in a given situation.} the number of spin- and color-states for initial state partons $i$ and $j$ and for the resonance $R$. In the narrow-width approximation, one focuses on the region $\hat{s} \approx m^2_R$ and approximates
\begin{equation}
\frac{1}
{(\hat{s}-m^2_R)^2 + m^2_R \Gamma^2_R}
\approx \frac{\pi}{m_R \Gamma_R} \delta(\hat{s} - m^2_R)~.
\end{equation}

\noindent Integrating over parton densities, and summing over incoming partons and over the outgoing partons which produce experimentally indistinguishable final states, we find the tree-level
hadronic cross section to be
\begin{align}
\sigma^{XY}_R &\ \equiv \sigma_R \times BR(R \to X + Y) = 16\pi^2 \cdot {\cal N} \cdot \frac{ \Gamma_R}{m_R} \times  \nonumber \\
& 
\left( \sum_{ij} (1 + \delta_{ij}) BR(R\to i+j) \left[\frac{1}{s} \frac{d L^{ij}}{d\tau}\right]_{\tau = \frac{m^2_R}{s}}\right) \cdot \left(\sum_{xy\, \in\, XY} BR(R\to x+y)\right)~. 
\label{eq:cross-section}
\end{align}
Here ${d L^{ij}}/{d\tau}$ corresponds\footnote{
In particular,
	\begin{equation}
	\left[ \frac{d{L}^{ij}}{d\tau}\right] \equiv 
	\frac{1}{1 + \delta_{ij}} \int_{\tau}^{1} \frac{dx}{x}
			\left[ f_i\left(x, \mu_F^2\right) f_j\left( \frac{\tau}{x}, \mu_F^2 \right) +
			f_{j}\left(x, \mu_F^2\right) f_i\left( \frac{\tau}{x}, \mu_F^2 \right) \right]  \,,
	\label{eq:lumi-fun}
	\end{equation}
	where in this paper, for the purposes of illustration, we calculate these parton luminosities using the {\tt CT14LO}~\cite{Pumplin:2002vw} parton density functions, setting the factorization scale $\mu_F^2= m^2_R$.} to the luminosity function for the $ij$ combination of partons and $X\, Y$ label the
	set of experimentally indistinguishable final states.

Defining a weighting function $\omega_{ij}$ allows us to reframe the sum over $ij$ as follows:
\begin{align}
\sum_{ij} & (1 + \delta_{ij}) BR(R\to i+j) \left[\frac{1}{s} \frac{d L^{ij}}{d\tau}\right]_{\tau = \frac{m^2_R}{s}} = 
\left[\sum_{ij} \omega_{ij} \left[\frac{1}{s} \frac{d L^{ij}}{d\tau}\right]_{\tau = \frac{m^2_R}{s}}\right] \cdot \left[\sum_{i'j'} (1 + \delta_{i'j'}) BR(R\to i'+j') \right] \nonumber 
\label{eq:rewritten}
\end{align}
where
\begin{equation}
\omega_{ij} \equiv \dfrac {(1 + \delta_{ij})  BR(R\to i+j)} {\sum_{i'j'} (1 + \delta_{i'j'}) BR(R\to i'+j')}~.
\end{equation}
The fraction $\omega_{ij}$ lies in the range $ 0 \le \omega_{ij} \le 1$ and by construction $\sum_{ij} \omega_{ij} = 1$.  

Substituting this into the cross-section in Eqn.~\ref{eq:cross-section}, we may obtain an expression for the product of the sums of incoming and outgoing branching ratios:
\begin{align}
\left[\sum_{i'j'} (1 + \delta_{i'j'}) BR(R\to i'+j') \right]  &\cdot \left(\sum_{xy\, \in\, XY} BR(R\to x+y)\right) = 
\label{eq:gen-bran-init}\\
&\frac{\sigma^{XY}_R} { 16\pi^2 \cdot {\cal N} \cdot \frac{\Gamma_R}{m_R} \times 
\left[\sum_{ij} \omega_{ij} \left[\frac{1}{s} \frac{d L^{ij}}{d\tau}\right]_{\tau = \frac{m^2_R}{s}}\right]} ~.\nonumber
\end{align}
This product is bounded from above by a value depending on the identities of the incoming ($i'j'$) and outgoing ($x,y$) partons,
\begin{equation}
BR(R\to i+j) (1 + \delta_{ij}) \cdot \sum_{xy\,\in\, XY} BR(R\to x+y)  \le
\begin{cases}
1/4 & i\neq j,\, ij \neq xy\in XY \\
1 & i\neq j,\, ij = xy\in XY \\
1/2 & i=j,\, x=y,\, ij\neq xy\in XY \\
2 & i=j, x=y, ij = xy\in XY
\end{cases}
\label{eq:combined}
\end{equation}
Framing the information in this way is what enables one to swiftly discern whether a given class of models is potentially consistent with a given data set.

Comparisons between data and theory are simplified by re-arranging Eqn.~\ref{eq:gen-bran-init}  so that the left-hand side includes the ratio of resonance width to mass. This defines the ``simplified limits variable'', $\zeta$:
\begin{align}
\zeta \equiv \left[\sum_{i'j'} (1 + \delta_{i'j'}) BR(R\to i'+j') \right]  &\cdot \left(\sum_{xy\, \in\, XY} BR(R\to x+y)\right) \cdot \frac{\Gamma_R}{m_R} = 
\label{eq:gen-bran-rat}
\\
&\frac{\sigma^{XY}_R} { 16\pi^2 \cdot {\cal N} \times 
\left[\sum_{ij} \omega_{ij} \left[\frac{1}{s} \frac{d L^{ij}}{d\tau}\right]_{\tau = \frac{m^2_R}{s}}\right]} ~. \nonumber
\end{align}
When working in the narrow width approximation, and assuming that $\Gamma/M \leq 10\%$, the upper bounds on the products of branching ratios mentioned above correspond to upper limits on $\zeta$ (Eqn. \ref{eq:combined}) that are a factor of ten smaller.

%%%%%%%%%%%%%%%%%%%%%%%%%%%%%%%%
% EXTENSION TO BROADER RESONANCES
%%%%%%%%%%%%%%%%%%%%%%%%%%%%%%%%

\section{Extending the Method to Broader Resonances}

We now generalize the results obtained in \cite{Chivukula:2016hvp} for resonances of larger widths by employing a Breit-Wigner representation of the resonance. We focus on resonances with fully-reconstructable final states: dileptons, dibosons, and dijets and on situations in which the resonance is far more massive than its decay products.

The total cross-section for the production and decay of an s-channel resonance in the channel $i + j \to R \to x + y$ can be obtained by convoluting the parton luminosity with the partonic cross-section $\hat{\sigma}(\hat{s})$ as follows

	\begin{equation}
	\sigma_R^{ij,xy} = \int_{s_{min}}^{s_{max}}d\hat{s}\,
	\hat{\sigma}(\hat{s}) \cdot \left[ \frac{d L^{ij}}{d\hat{s}}\right]~, {\rm where}
	\label{eq:BW-cs}
	\end{equation}
	\begin{equation}
	\hat{\sigma}(\hat{s})^{ij,xy} \equiv  \frac{\Gamma^2_R}{m^2_R} \cdot\frac{\hat{s}}{m^4_R}\cdot
	\frac{16 \pi{\cal N} (1 + \delta_{ij}) BR(R\to i+j) \cdot BR(R\to x+y)}
	{\left(\frac{\hat{s}}{m^2_R}-1\right)^2+\frac{\Gamma^2_R}{m^2_R}}~.
	\label{eq:sigshat}
	\end{equation}
One can parametrize the cross-section in eqn.~\ref{eq:sigshat} in terms of the resonance mass ($m_R$), its width-to-mass ratio ($\Gamma_R/m_R)$, and the product of the relevant branching ratios ($BR_{ij}\cdot BR_{xy})$.

In arriving at the form of eqn.~\ref{eq:sigshat}, we have made several approximations.  Because we are studying systems where a heavy resonance is decaying to states that are far lighter than $m_R$, we have approximated each of the running partial-widths $(\Gamma(\hat{s}))$ of the resonance in the numerator by a phase space factor times the on-shell partial-width $\sqrt{\hat{s}} \Gamma / m_R$; this gives rise to the factor of $\hat{s}/ m_R$ that impacts the overall magnitude of $\hat{\sigma}(\hat{s})$.   Corrections to this approximation are suppressed by powers of $m^2/\hat{s}$, where $m$ is the mass of standard-model particles in the decay -- and are therefore negligible for TeV scale resonance searches.
At the same time, we noted that the presence of the running total-width in the denominator serves mainly to shift the location of the resonance peak; we have neglected this smaller effect, replacing  $\sqrt{\hat{s}}\Gamma_R(\hat{s})$ by $m_R \Gamma_R$.  We have checked numerically that this second approximation has a negligible effect unless $\Gamma_R \simeq m_R$.

In general, experiments searching for resonances present their constraints either as limits on $\sigma \times BR$ or in terms of parameters of a given model. 
More specifically, experiments count the number of events in each bin of the invariant mass distribution.  Constraints are set by defining likelihoods (usually Poissonian) for signal and signal + background hypotheses and performing statistical tests on these hypotheses. 
The theoretical prediction for signal cross-section is determined in terms of a model. 
Our proposal of simplified limits simply replaces the theoretical model prediction with expressions of the cross-sections given above. Now instead of couplings and masses, constraints are placed on  the parameters of simlified limits-- $\zeta$ and the mass of the resonance. In order to map simplified limits to a specific model one would need to specify $\omega_{ij}$ as well.
In this work, we do not follow the procedure described above to extract limits, simply because the full likelihoods, nuisance parameters and errors are not available. We instead use ``Brazil band" plots provided by experimental papers to extract the $2\sigma$ exclusion on $\sigma\times BR$ as a function of mass. We then equate the extracted cross-section to the expressions for cross-section given earlier in order to extract $2\sigma$ constraints on the parameter $\zeta$. This provides only an approximate estimate of the exclusion on $\zeta$, which suffices for our purpose of demonstrating the salient features of simplified limits. We take care to integrate $\hat{s}$ in eqn.~\ref{eq:sigshat} only over the range specified by the kinematic cuts implemented in each experiment.

\subsection{Extensions and Limitations of this Approach}
\label{subsec:limitations}

As mentioned above, the simplified limits approach provides a compact and easily interpretable method of presenting limits on resonance searches. In this section we list possible extensions of this approach as well as describing some limitations.
\begin{itemize}
	\item  As is done in traditional limits on cross-section times branching ratio, it is possible to include higher order corrections to limits on $\zeta$ by simply using K-factors. However, one has to be careful when higher order effects change the acceptance due to kinematical cuts. 
	\item The acceptance also depends on the spin of the resonance. As is sometimes done in the case of traditional Brazil band plots where limits are displayed in terms of $\sigma\times\text{Branching Ratio}\times\text{Acceptance}$, one could also present limits in terms of $\zeta\times\text{Acceptance}$.
	\item The simplified limits approach is most directly applicable to searches where the kinematics of the final state can be reconstructed entirely, i.e. when searching for bumps in a invariant mass spectrum. For resonance searches in which the invariant mass spectrum cannot be reconstructed, such as ($W^{\prime}\to l \nu$), other kinematic variables (transverse mass for $W^{\prime}\to l \nu$ ) are analyzed. One could also apply the simplified limits approach in this case. However, again one needs to be careful about issues of acceptance and kinematic cuts.
	\item The simplified limits approach works when interference of the BSM production process with SM backgrounds is negligible. This is the case for most $s$-channel resonance processes.
		\item Simplified limits for resonances produced in pairs or produced in association with other particles may be interetsting to consider in the future.

		\end{itemize}
	
Keeping in mind the limitations of our method, we restrict our attention to $s$-channel resonance searches in which the invariant mass of the resonance can be completly reconstructed, and consider leading-order analyses. 

%%%%%%%%%%%%%%%%%%%%%%%%%%%%%%%%
%  APPLICATIONS TO BROADER RESONANCES
%%%%%%%%%%%%%%%%%%%%%%%%%%%%%%%%

\section{Applications to Broader Resonances}

We will now apply the extended simplified limits technique to various situations of general theoretical and experimental interest.  Here we discuss electrically neutral spin-1 resonances decaying to dileptons; a $W'$ state decaying to dibosons; and resonances of various spins and colors decaying to dijets.  Note that in each of these examples the final state may be fully reconstructed.

In the first example, we will illustrate the power of the extended simplified limits analysis by showing results both in terms of an upper bound on the resonance's combined production and decay branching ratios and  separately in terms of a bound on the simplified limits variable $\zeta$.  Thereafter, we will show results only in terms of $\zeta$.  In discussing each application, we will show how the limits compare when the resonance is treated in the narrow width approximation (NWA) or assumed to be broader and treated as a Breit-Wigner shape (BW).  

Throughout, we will show the observed limits on $\zeta$ corresponding to LHC data from the ATLAS or CMS experiments, and in some cases we also show the expected limits.  As discussed in  \cite{Chivukula:2016hvp} if the observed limit is ever seen to be much weaker than the expected limit, meaning that some evidence of a new state has been found, then a given class of resonance (with a particular set of dominant production and decay modes) will be a candidate explanation for the excess only if the product of branching ratios (or corresponding value of $\zeta$) required to produce the observed signal falls in the physical zone (e.g., the product of branching ratios can never be required to exceed 1).

In each application, we separately illustrate the specific value the $\zeta$ variable takes in benchmark theoretical models from the literature.  Again, as discussed in \cite{Chivukula:2016hvp},  if an excess were found, only a model whose predicted value of $\zeta$ fell in the window between the expected and observed limits on $\zeta$ would be a good candidate for explaining the excess.

%%%--------------------------------------------------
%%%        Example: Dileptons      
%%%--------------------------------------------------

\subsection{ $u\bar{u} + d\bar{d}\to R \to l\bar{l}$}
\label{app:dileptons}

In this application, we study colorless spin-1 resonances that decay to dileptons.  We employ the ATLAS analysis~\cite{ATLAS-CONF-2016-045} of dilepton final states at $\sqrt{s}=13 ~\text{TeV}$ as the source of our information on the observed limits on branching ratios or $\zeta$.  The cuts used to identify events for this analysis are summarized in  Appendix A for the reader's convenience.

In order to extract the $\zeta$ variable we 
assume that the acceptance times efficiency for the resonances under consideration would be identical to that of the $Z^{\prime}$ considered by the ATLAS experiment. 
Since the only kinematic cuts employed are those on rapidity and transverse momentum, the geometrical  acceptance depends only on the spin of the resonance -- in this case a spin-1 resonance. In the dijet applications discussed later on, we will study resonances of different spins and our analysis will specifically incorporate the impact of resonance spin upon acceptance.

\begin{figure}[h]
	\includegraphics[width=0.65 \textwidth]{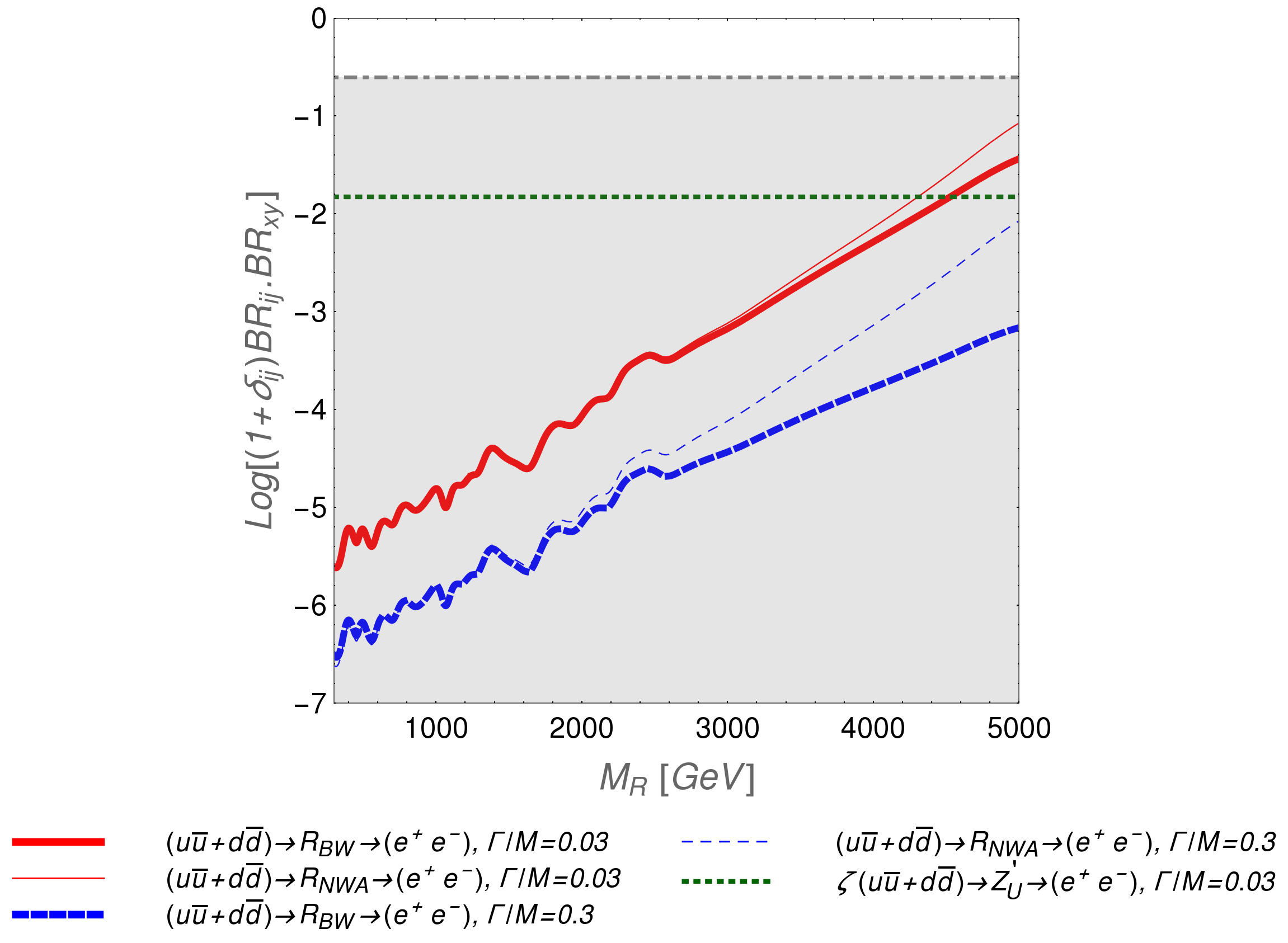} 
	\includegraphics[width=0.65 \textwidth]{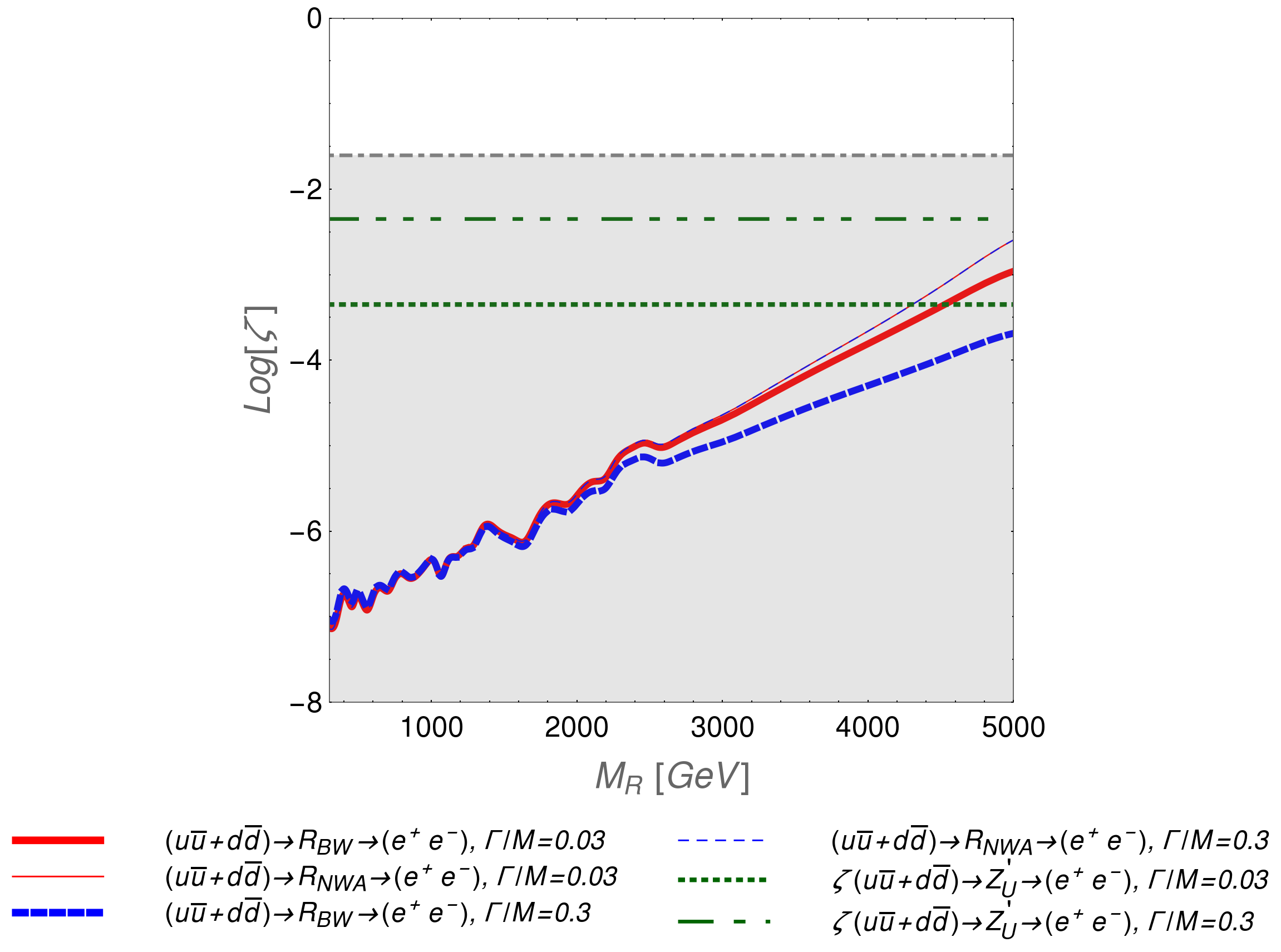}
		\caption{ \small \baselineskip=3pt  Simplified Limits on a flavor-universal spin-1 resonance decaying to dileptons, shown in terms of branching ratios (above) and $\zeta$ (below).  In both panes, the diagonal curves show the observed experimental ATLAS~\cite{ATLAS-CONF-2016-045} upper limits (at 95\% credibility level) on a $Z'$ boson decaying to dileptons.  The thin curves treat the resonance in the narrow width approximation (NWA); the bold curves are for a finite-width resonance in the Breit-Wigner (BW) approximation.  The shaded rectangle in each pane is the area in which the product of branching ratios (or $\zeta$) has a physically reasonable value (see text); the horizontal green dotted curve shows the theoretical value of the vertical axis variable corresponding to a benchmark flavor-universal $Z^{\prime}$ boson -- and the excluded region for that model coresponds to masses below the intersection of that line with the corresponding diagonal curve.  {\textbf{Above: }} Observed upper limit on the product of branching ratios as a function of resonance mass. The upper solid red pair of curves is for $\Gamma/M = 0.03$; the lower dashed blue pair is for $\Gamma/M = 0.3$.  In each case, using the NWA gives a conservative upper limit, in the sense of excluding a somewhat smaller region of masses.  {\textbf{Below: }} The same upper limits, re-expressed in terms of the simplified-limits variable $\zeta$. The thin NWA curves now overlap since the value of $\Gamma/M$ is incorporated within $\zeta$; the bold BW curves for  resonances of different widths are distinct.  Note that a broader BW resonance pulls away from the NWA curve at a relatively lower value of $M$. (Note that, by definition, the bounds on $\zeta$ in the NWA are independent of $\Gamma/M$.) }
		\label{fig:simlim-zprime}
\end{figure}

Fig.~\ref{fig:simlim-zprime} shows the observed upper limits\footnote{Differences between the limits as displayed here and as originally reported in ref.~\cite{ATLAS-CONF-2016-045} arise from choice of PDF (and scale) and the use of a mass dependent K-factor ($\sim \in [1.1,1.3]$).} (at 95\% credibility level)  on hadronically-produced vector resonances decaying to dielectron final states, expressed through the simplified limits analysis.
The upper pane of Fig.~\ref{fig:simlim-zprime} shows upper limits on the value of the product of branching ratios $BR(j\bar{j})BR(e^+e^-)$, where $j=\{u,d\}$. Here we have assumed universal couplings to quarks (as with a resonance coupling to baryon number) and neglected the small contribution of $(s,c,b)$ quarks to the resonance production cross-section. Similarly the lower plot of Fig.~\ref{fig:simlim-zprime} shows upper limits on $\zeta$.
The thicker lines correspond to using the Breit-Wigner (BW) distribution to evaluate upper limits whereas the thinner lines are evaluated using the narrow width approximation (NWA).
The grey-shaded rectangle in the upper pane is the area in which the product of branching ratios is physical: it cannot exceed 1/4 since the initial and final states are different, and neither the two initial nor the two final state particles are identical. The grey-shaded rectangle in the lower pane is the corresponding physical region of $\zeta$, given that we are assuming $\Gamma_R / M_R < 0.3$ in our BW analysis.  

From examining either pane, we can see that using the narrow width approximation gives a conservative upper limit on the vertical-axis variable, in the sense of not overstating the strength of the bound.  In the upper pane, the upper bounds on resonances of $\Gamma/M = 0.3$ and $\Gamma/M = 0.03$ are vertically displaced from one another by an order of magnitude.  When the same upper limits are re-expressed in terms of the simplified-limits variable $\zeta$ in the lower pane, the thin NWA curves now overlap since the value of $\Gamma/M$ is incorporated within $\zeta$. The bold curves for Breit-Wigner resonances of different widths are distinct and we can observe that a broader BW resonance pulls away from the NWA curve at a relatively lower value of $M$.

We use as our comparison a benchmark $Z'$ model which couples universally\footnote{In general, the $U(1)$ gauge theory of a $Z'$ coupling universally to all quarks would have gauge and/or gravitational anomalies and a full model could require additional spectators.   We use this object here purely as an illustration.} to all quarks (in particular, which has the same value of $g^2_L+g^2_R$ for all up- and down-quarks); a $Z'$ coupling to $B-L$ would be a familiar example of such a $Z'$. The horizontal green dotted lines in the upper (lower) pane correspond to the product of branching ratios (value of $\zeta$) for a resonance of this sort and the indicated values of the resonance width-to-mass ratio.  From either pane, it is clear that the ATLAS upper limits on the vertical-axis variable exclude this particular benchmark model for $Z'$  masses below at least 4 TeV.

Limits that are set using the BW shape tend to be stronger than those set using the NWA, expecially at larger masses. In other words, the cross-section as evaluated using the NWA is smaller than the cross-section as evaluated using the BW shape.
	This occurs because large mass resonances require a large parton momentum fraction ($x$) to be produced. At large values of $x$, the parton distribution functions fall rapidly. The BW resonance integrates some of the luminosity $\sqrt{\hat{s}} < M$, thus giving rise to a larger cross-section.
	So long as there are no additional kinematic cuts (especially those affecting the invariant mass distribution, see Sec. \ref{app:dibosons}), this pattern is typical of the limits set on resonances.

\begin{figure}[h]
	\includegraphics[width=0.7 \textwidth]{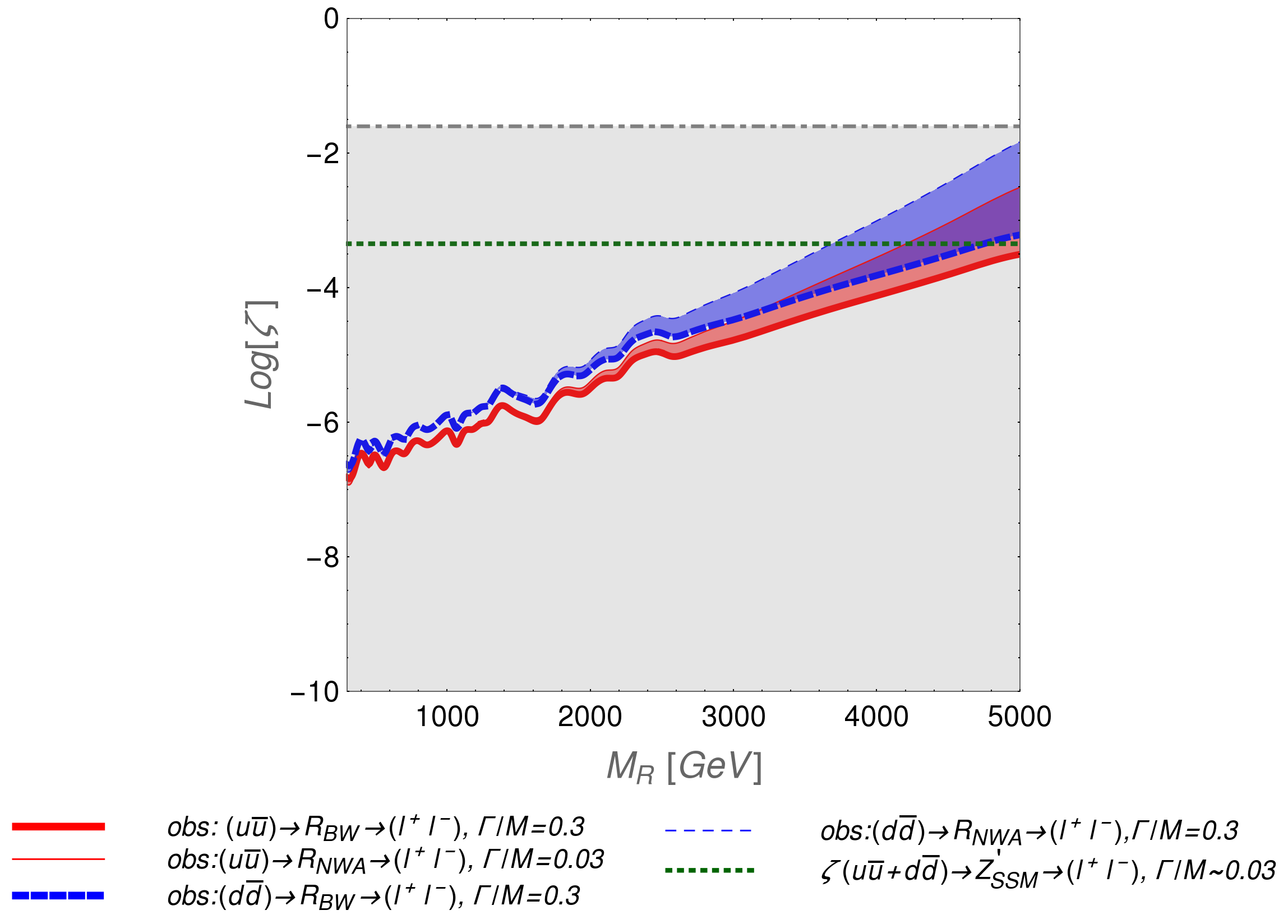}
		\caption{ \small \baselineskip=3pt  Simplified Limits on a spin-1 resonance decaying to dileptons and coupling either only to down-type or only to up-type quarks. The diagonal curves show the observed ATLAS \cite{Aad:2015owa} upper limits (at 95\% confidence level) on a vector boson decaying to dileptons, displayed in the $M$ vs. $\zeta$ plane. The shaded region is the area in which $\zeta$ has a physically reasonable value (see text) and the nearly-horizontal dotted green line shows the theoretical value of the vertical axis variable corresponding to a SSM $Z'$ boson for comparison.  The various diagonal curves compare several analyses.  The upper blue pair (lower red pair) of curves represent the bounds on a Z' boson that couples only to and is produced only by down-type (up-type) quarks.  The limit on any particular vector resonance coupling to quarks will lie between these extremes.  Within each pair, the upper, thin curve was derived in the narrow width approximation while the lower thick one was derived assuming the resonance was a Breit Wigner form with $\Gamma/M = 0.3$; the shading between the members of a pair highlights the difference between the limit in the NWA and BW approximations. }
		\label{fig:simlim-zprime-uudd}
\end{figure}

While Fig.~\ref{fig:simlim-zprime} was produced under the simplifying assumption that the resonance had flavor-universal couplings to quarks, that assumption does not hold for most models.  Since the relative strength of a resonance's couplings to $u\bar{u}$ and $d\bar{d}$ will affect its production cross-section, we have also made a more general analysis.  Fig.~\ref{fig:simlim-zprime-uudd} illustrates the degree to which the relative strength of the couplings to up-type and down-type quarks impacts the simplified limits on spin-1 resonances decaying to dileptons. Again, the grey-shaded rectangle shows the physical region of $\zeta$, given that we take $\Gamma_R / M_R < 0.3$ in our BW analysis.   The upper pair of diagonal curves represent the 95\% confidence level upper bounds on $\zeta$ for a vector boson that couples only to down-type quarks; the upper, thin bound curve was derived in the NWA while the lower thick one was derived assuming the resonance has a Breit-Wigner form with $\Gamma/M = 0.3$.  The lower pair is similar, but derived under the assumption that the vector boson couples only to up-type quarks.  The limit on any intermediate case where the resonance couples to both up-type and down-type quarks will lie between these extremes; the difference in the strength of the bound on $\zeta$ varies from a factor of a few at low $M_R$ to nearly a factor of ten at high $M_R$.  

As a benchmark, the horizontal dotted line shows the value of $\zeta$ for a Sequential Standard Model $Z^{\prime}_{SSM}$ boson, which, like the Standard Model $Z$ boson, has unequal couplings to up-type and down-type quarks.  The ATLAS upper limits on $\zeta$ exclude this benchmark model for resonances masses below at least 4 TeV.

%%%--------------------------------------------------
%%%        Example: WZ     
%%%--------------------------------------------------

\subsection{Example: $ud \to R \to  W^{\pm} Z$ }
\label{app:dibosons}

In this second application, we study colorless, electrically-charged spin-1 resonances that decay to $WZ$.  
We use the ATLAS analysis~\cite{ATLAS-CONF-2016-055} with $ 15.5\invfb$ at $\sqrt{s} = 13$~TeV as the basis for our work.  In this analysis ATLAS searched for resonances with mass $M_{R}> 1$~TeV decaying to dibosons ($WW$, $WZ$, $ZZ$ ), in the fully hadronic channel $qqqq$.  Selection criteria are summarized in Appendix B for the reader's convenience.

\begin{figure}[h]
	\includegraphics[width=0.54 \textwidth]{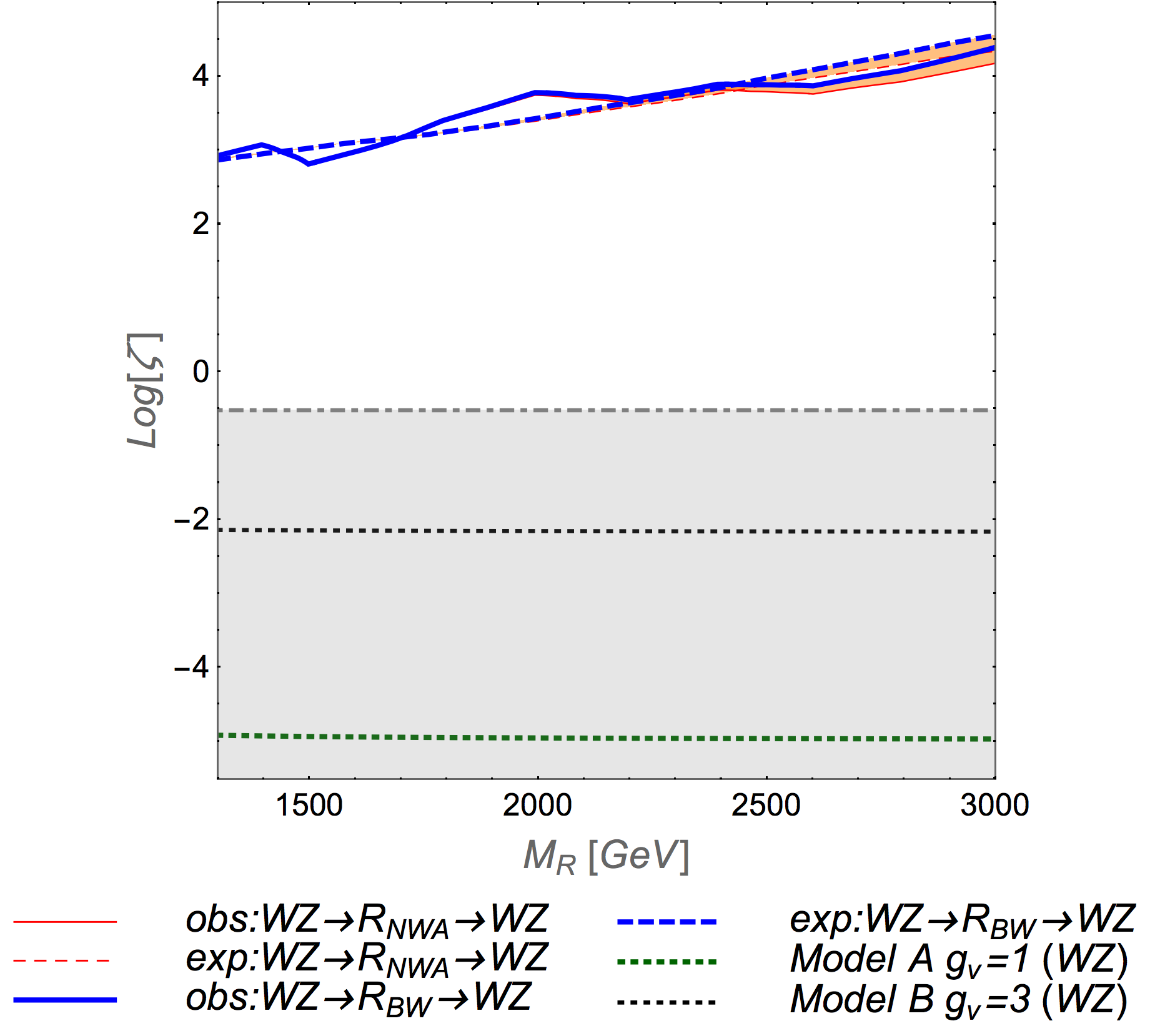} \qquad\quad
		\includegraphics[width=0.61 \textwidth]{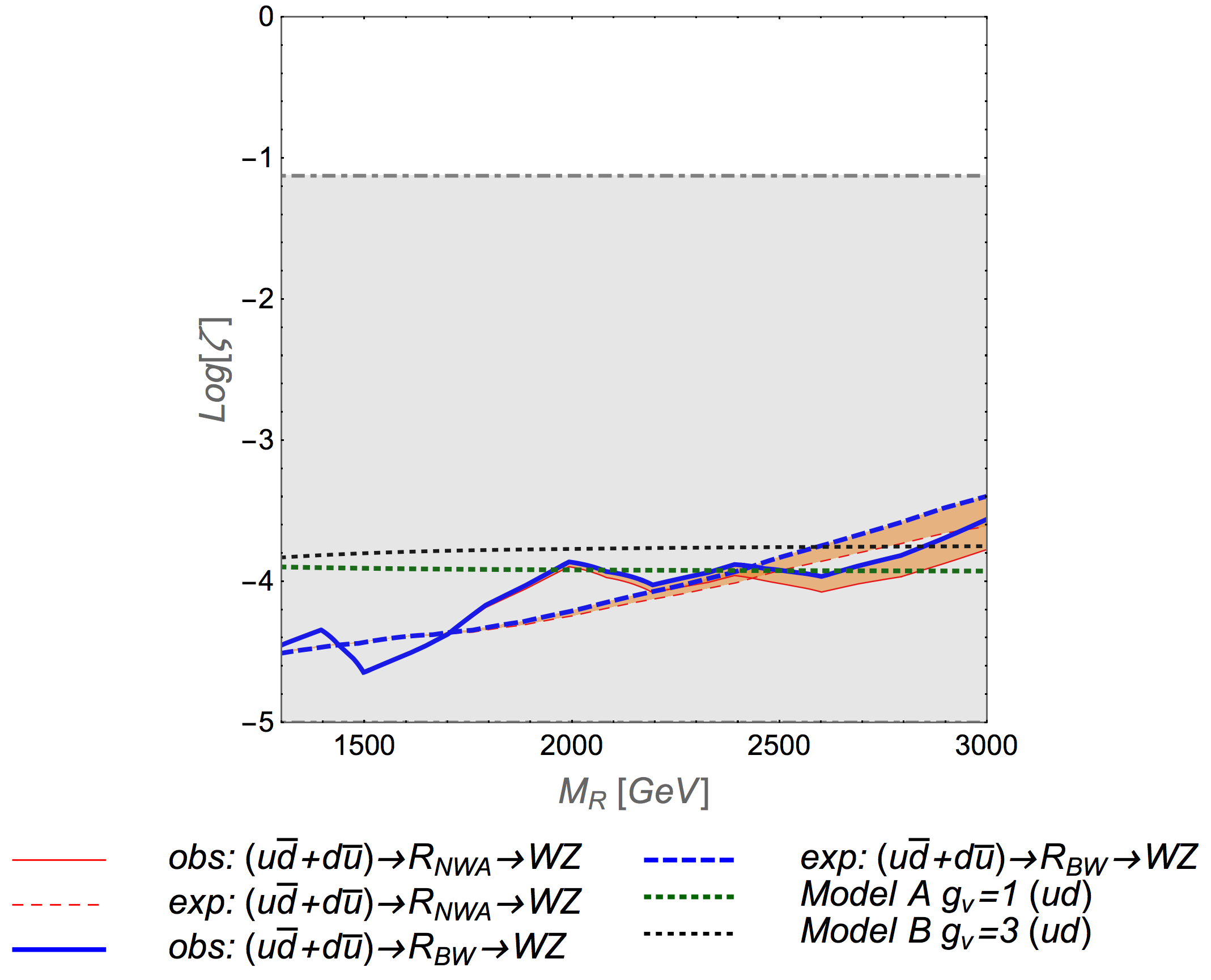}

			\caption{ 
				\small \baselineskip=3pt  Simplified Limits on a spin-1 resonance decaying to dibosons.  The diagonal curves show the expected (dashed) and observed (solid) ATLAS \cite{ATLAS-CONF-2016-055} upper limits (at 95\% confidence level) on a W' boson decaying to $W^{\pm}Z$ final states, in the $M$ vs. $\zeta$ plane. Within the shaded rectangles, $\zeta$ has a physically reasonable value  (see text). The nearly-horizontal dotted lines show the theoretical value of $\zeta$ in the benchmark Heavy Vector Triplet models of ~\cite{Pappadopulo:2014qza}.  {\textbf{Upper: }} The case of a vector resonance produced by $WZ$ fusion.  Thinner curves were derived using the NWA; thicker curves were derived assuming the resonance has a Breit Wigner form with $\Gamma/M = 0.3$. The gold-shaded region between the NWA and BW curves for a given kind of limit (observed or expected) illustrate the degree to which the NWA and BW approximations yield different results.  As discussed in the text, the BW gives a slightly more conservative limit here, due to the impact of cuts. Note the value of $\zeta$ required to conform with observation lies several orders of magnitude outside the shaded region. {\textbf{Lower: }} The case of a vector resonance produced through quark/anti-quark annihilation.  The conventions remain the same. Note that the upper bounds on $\zeta$ lie well within the shaded region. }
				\label{fig:simlim-Wprime}
\end{figure}

Fig.~\ref{fig:simlim-Wprime} shows the expected and observed 95\% confidence level limits on $\zeta$ for vector resonances decaying to diboson final states.  As above, the grey-shaded region is the area in which $\zeta$ has a physically reasonable value, given that we are assuming $\Gamma_R / M_R < 0.3$ in our BW analysis; in the upper pane where the initial and final states are identical, the product of branching ratios is bounded from above by 1, while in the lower pane it is bounded from above by 1/4.   In both panes, the diagonal solid curves correspond to observed limits while the diagonal dashed ones correspond to expected limits.  The thinner red curves have been derived using the NWA and the thicker blue ones have been derived assuming a Breit-Wigner form for the resonance, with $\Gamma/M = 0.3$.   Also shown in each pane are two horizontal short-dashed curves corresponding to the value of $\zeta$, as a function of resonance mass, for our comparison benchmark models: the Heavy Vector Triplet (HVT) Models \cite{Pappadopulo:2014qza}.  The HVT phenomenological model was introduced to study charged vector bosons potentially coupling both to fermions and to electroweak bosons, and following \cite{Pappadopulo:2014qza} we illustrate with two choices for the defining paramter, set A ($g_V=1$; bold) and set B ($g_V=3$; thin).

The upper pane explores the situation where the vector resonance is both produced through $WZ$ fusion and decays back to a hadronically-decaying $WZ$ pair.  We see that the upper limit on $\zeta$ lies orders of magnitude outside the (shaded) region of physically reasonable values of $\zeta$.  One implication is that the upper bound is too weak to give a meaningful constraint on a fermiophobic vector resonance. Another is that a fermiophobic vector resonance would not be a viable candidate to explain any signs of an excess of events;  e.g., if one were to interpret the fact that the observed limit lies above the expected limit near $M_R = 2$ TeV as possible evidence of a resonance, one would have to look to another class of resonance for an explanation.  All of this is consistent with the results from \cite{Chivukula:2016hvp}.  

In contrast, the lower pane explores the case where the vector resonance is produced through $u\bar{d} + d\bar{u}$ initial states and decays to $WZ$; here, the observed upper bound lies well within the physical region.  So if the area around $M_R = 2$ TeV (where the observed limit is weaker than expected) were taken as a possible locus of a new resonance, this production mode would be a viable candidate. In addition, we see that the $\zeta$ values predicted by Models A and B both fall within or near the ``window'' between observed and expected limits, which would make them worthy of further examination.  Again, this is consistent with Ref. \cite{Chivukula:2016hvp}.

What is new here is that we can see the impact of going beyond the narrow width approximation.  As shown in Fig.~\ref{fig:simlim-Wprime}, the limits obtained from assuming a BW distribution are similar to those obtained in the NWA, but not identical.  In this case, ATLAS selected events such that the invariant mass of the two-fat-jet system lies in the range $1.0~\text{TeV}< m_{JJ}< 3.5~\text{TeV}$. The presence of the hard upper bound on the invariant mass results tends to ``clip" the high-mass end of the broader BW signal distribution in a way that does not happen for the NWA case (where all signal events are in a single invariant mass bin).  As a result, the upper bound on $\zeta$ turns out to be slightly weaker than the limit derived using the  NWA limit, contrary to what we observed in the dilepton example.  However, just as in the dilepton example, the NWA limit still gives a solid first estimate of the simplified limits constraint even for a resonance of moderate width.

%%%--------------------------------------------------
%%%        Example: Dijets    
%%%--------------------------------------------------

\subsection{Example: Dijets }
\label{app:dijets}

We now apply the extended framework to new resonances that decay to dijets. We use the CMS results on dijets with $20\invfb$ of $8$~TeV data~\cite{Khachatryan:2015sja} as the source of our limits on $\zeta$; the various selection criteria are summarized in Appendix C for the reader's convenience. We study a variety of scenarios, including scalar, vector, and spin-2 resonances, for interpreting the experimental data.  These cases not only show how the resonance's spin impacts the bounds but also illustrate how the limits on resonances having the same spin but being produced through different initial-state partons can differ, due to the impact of the parton distribution functions.  Appendix C describes how we account for the impact of resonance spin on detector acceptance.

Alongside the experimental limits on $\zeta$ for each scenario, we show the predicted $\zeta(M_R)$ for one or two benchmark models from the literature.  Our benchmarks for scalar resonances decaying to dijets are the scalar octet resonance of  \cite{Hill:1991at,Frampton:1987dn,Martynov:2009en,Bai:2010dj,Harris:2011bh} and the scalar diquark \cite{Angelopoulos:1986uq,Hewett:1988xc,King:2005jy,Kang:2007ib}; the vector resonances we use as benchmarks are the Sequential Standard Model $Z^{\prime}$ and flavor-universal Colorons \cite{Frampton:1987dn,Frampton:1987ut,Chivukula:1996yr,Simmons:1996fz}.  We use the excited quarks from \cite{Baur:1989kv,Baur:1987ga,Redi:2013eaa} and the RS Graviton \cite{Randall:1999ee,Bijnens:2001gh} as samples of fermionic and spin-2 resonances, respectively.
 
\begin{figure}[h]
	\includegraphics[width = 0.51 \textwidth]{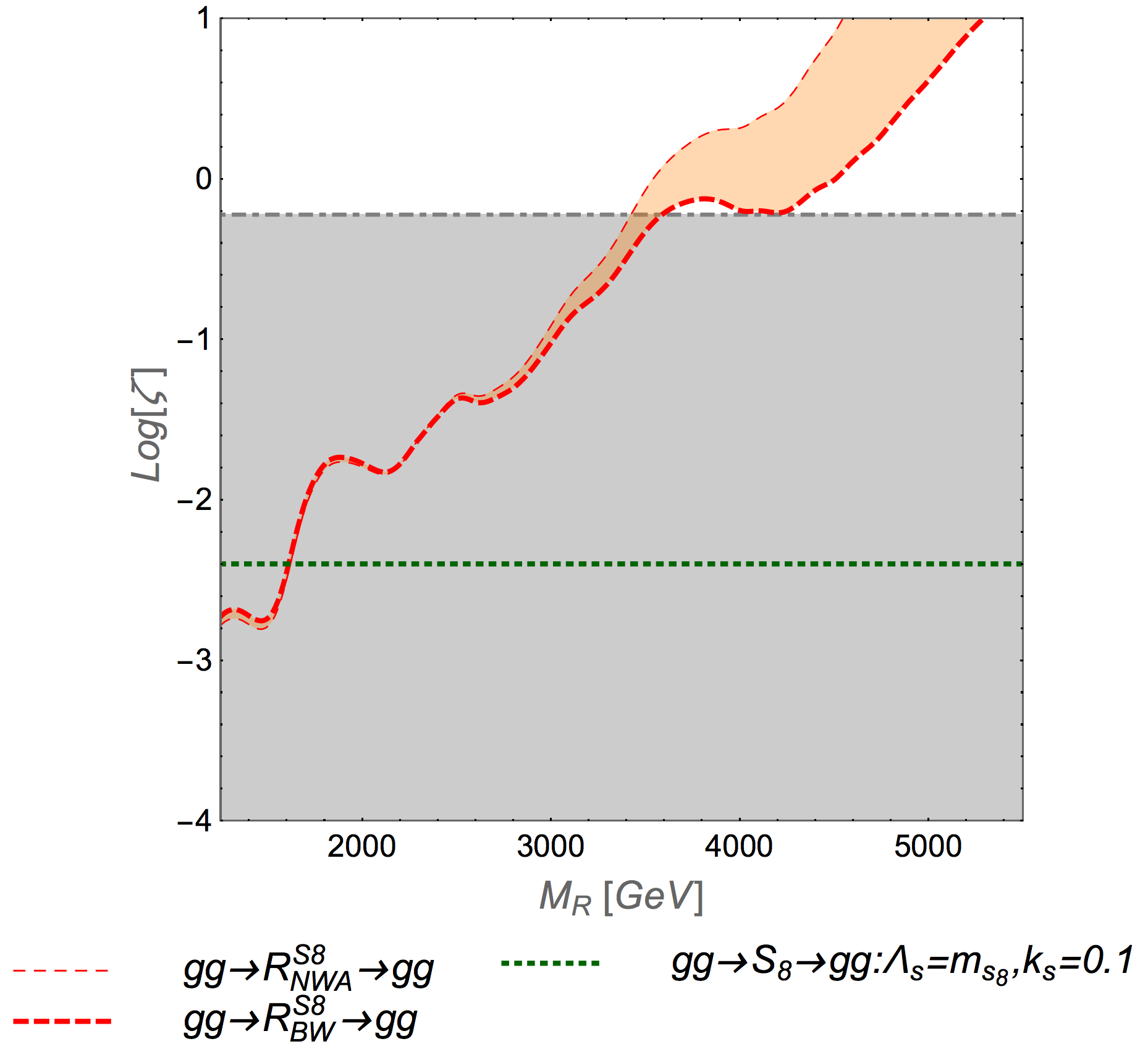} \quad\qquad
	\includegraphics[width = 0.57 \textwidth]{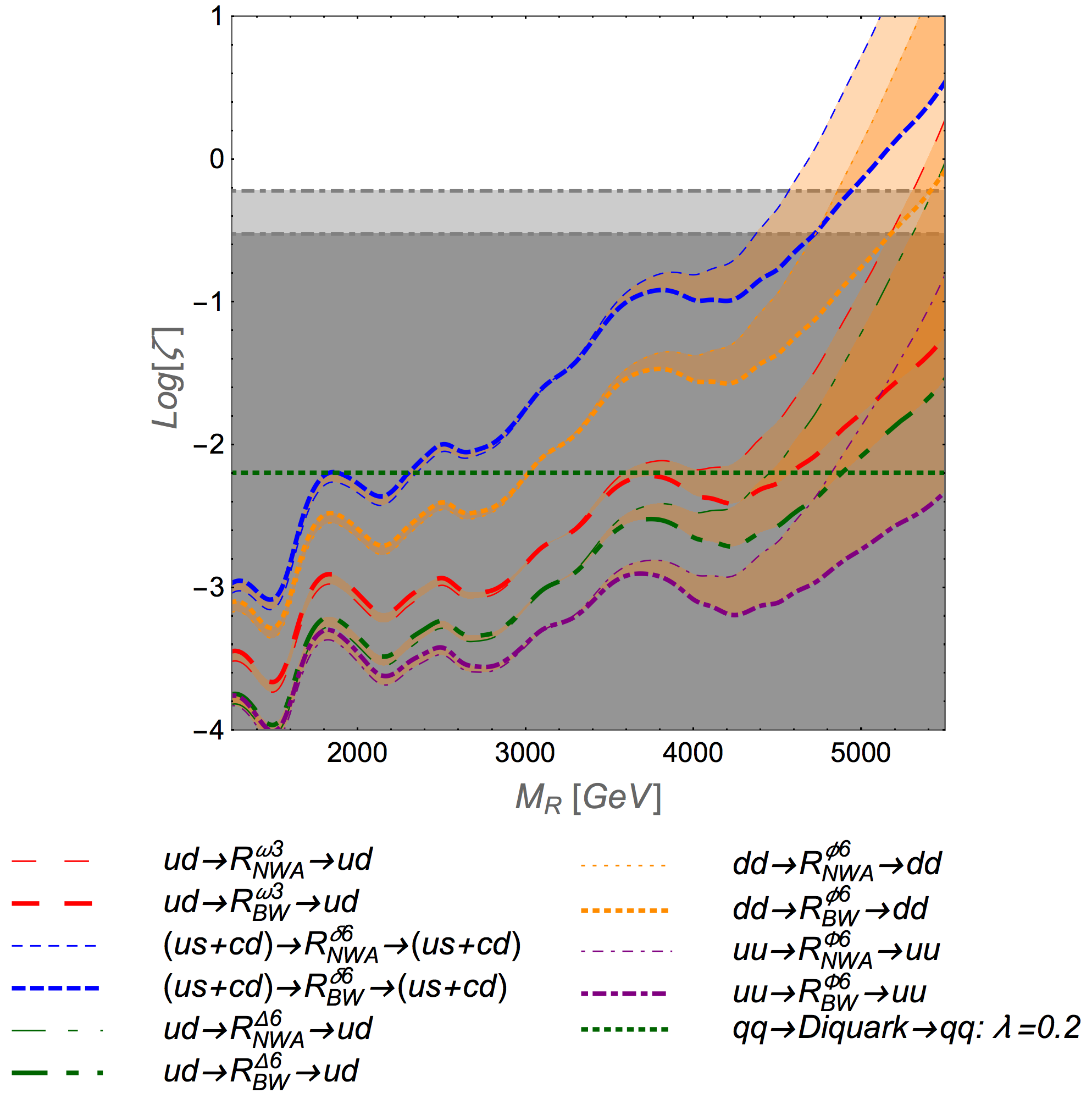}
	\caption{ \small \baselineskip=3pt Observed upper bounds (at 95\% confidence level) on scalar resonances from CMS dijet resonance searches ~\cite{Khachatryan:2015sja} shown in the $\zeta - M_R$ plane. Within the shaded rectangles, $\zeta$ has a physically reasonable value (see text). {\bf Upper:} The case of a scalar octet resonance produced by gluon fusion and decaying to gluon pairs. The gold-shaded region between the thin dashed diagonal NWA curve and the thick dashed diagonal BW curve (with $\Gamma/M = 0.3$) indicates the difference between the results in those approximations.  We find the NWA yields a conservative upper limit on $\zeta$.  The horizontal green dotted line shows the value of $\zeta$  for the benchmark scalar octet model \cite{Hill:1991at,Frampton:1987dn,Martynov:2009en,Bai:2010dj,Harris:2011bh}. {\bf Lower:} The cases of several different scalar diquark resonances $(\omega_3,\delta_6,\Delta_6,\phi_6,\Phi_6)$, as in the text.  The conventions for thin (NWA) and thick (BW) diagonal curves and the gold-shaded regions between them are as above.  Note the NWA and BW curves are indistinguishable for resonance masses below about 3.5 TeV; at higher masses, the NWA gives a conservative upper limit on the value of $\zeta$. The horizontal green dotted line marks the value of $\zeta$ common to all these diquark models, with a coupling $\lambda=0.2$.  }
	\label{fig:scalar-dijet}
\end{figure}

In Fig. \ref{fig:scalar-dijet} we consider the 95\% confidence level upper limits on color-octet scalar resonances produced via gluon fusion (upper panel) and on scalar diquarks produced by quark fusion (lower panel). In the upper panel, the shaded rectangle encompasses the area corresponding to physically reasonable values of $\zeta$, as understood via Eq. (\ref{eq:combined}) with $\Gamma_R / M_R \leq 0.3$ to be 0.3. The gold shaded region illustrates the difference between using the narrow width and Breit-Wigner approximations.  In the lower panel, the dark-shaded rectangle applies to all diquarks and the light-shaded extension applies only to cases with identical incoming partons ($uu$ or $dd$).  This panel highlights the dramatic difference in the ranges of model parameters excluded depending on the flavor properties of the diquark - and hence the flavor composition of the incoming partons; accordingly, we obtain lower mass limits from less than 2 TeV to more than 5 TeV, for the benchmark model illustrated. 

For scalars decaying to dijets, we find that using the NWA somewhat understates the LHC reach; again, that approximation therefore provides a conservative upper limit on the value of $\zeta$.  For the color-octet scalar, the NWA and BW curves are quite close together except at the higher resonance mass values where the experimental constraints also become too weak to impact the physical region of $\zeta$.  For diquarks, the experimental limits generally fall within physical region for $\zeta$, so that the divergence between the BW and NWA curves, including the larger separation at high mass values, is potentially of greater importance.

When we compare the results in Fig. \ref{fig:scalar-dijet} with those for a vector resonance decaying to dileptons in Fig. \ref{fig:simlim-zprime-uudd}, we see that the BW curve begins to visibly diverge from the NWA curve at different resonance masses: 1.5 TeV for a colorless vector resonance produced through $u\bar{u}$ or $d\bar{d}$ annihilation, 2.5 TeV for color octet scalars produced via $gg$ fusion, and 3.5 TeV for diquarks produced from $uu$ or $dd$ .  However, due to the properties of the parton luminosity functions of the incoming states, the BW curve falls below the NWA curve more rapidly for the dijet scenarios; in all three cases, the NWA and BW curves are an order of magnitude apart in $\zeta$ for a resonance mass of 5 TeV.

\begin{figure}[b]
	\includegraphics[width = 0.45 \textwidth]{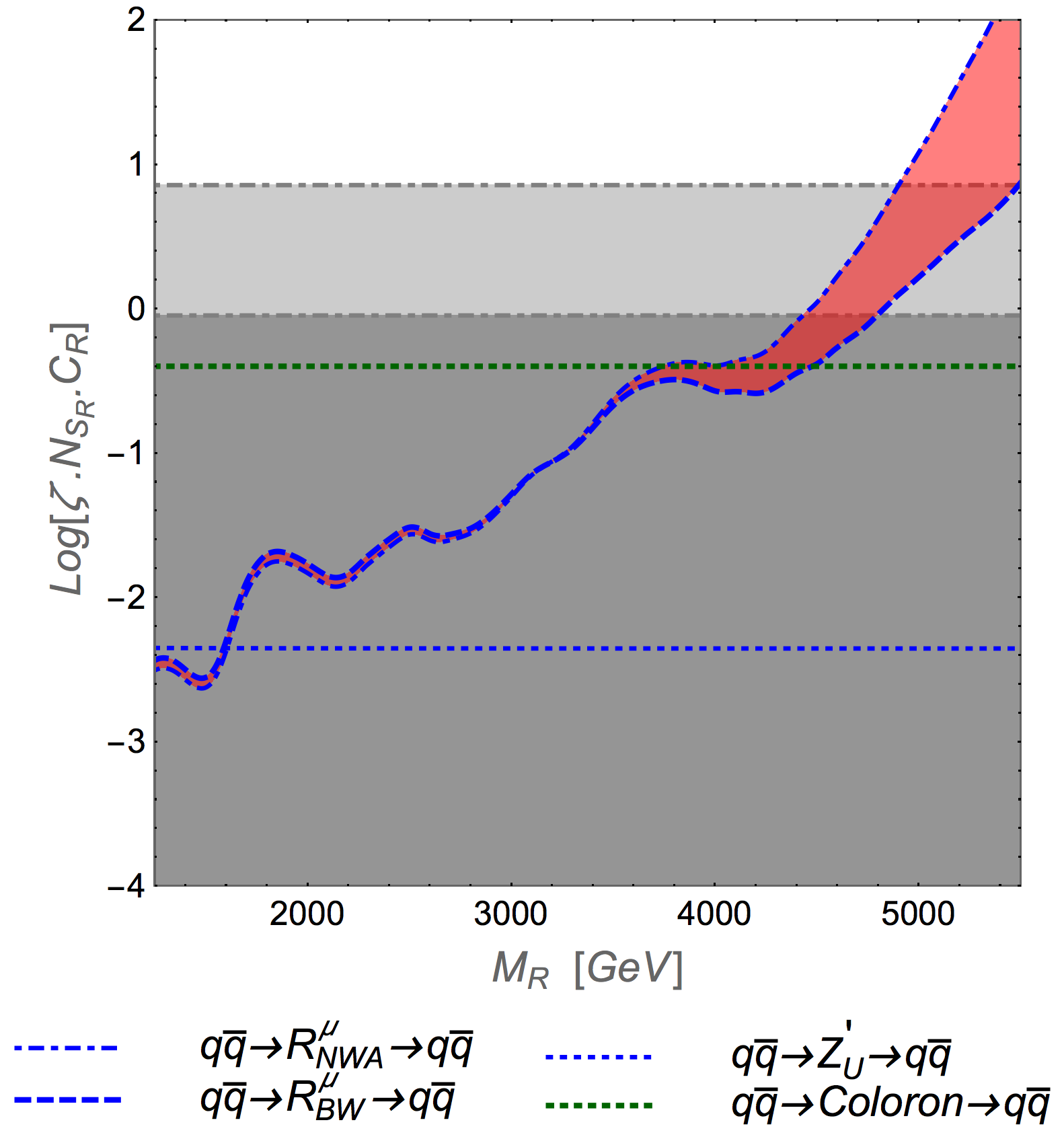}
	\caption{ \small \baselineskip=3pt Observed upper bounds (at 95\% confidence level) on flavor-universal vector resonances from CMS dijet resonance searches \cite{Khachatryan:2015sja} in the $M_R$ vs $\zeta N_{S_R} C_R$ plane.  Here, we have incorporated the factor $N_{S_R} C_R$ from Eq.~\ref{eq:nr} explicitly within the vertical axis variable; as a result, the limits on flavor universal vector resonances produced only via quark/anti-quark annihilation overlap one another. The impact of a finite width for the resonances is apparent from the red-shaded gap between the NWA and BW curves at high resonance mass; note that the BW limit (thick dashed line) including finite width is stronger than that obtained in the NWA (thin dashed line).  The dark-shaded rectangle shows the physical region of $\zeta N_{S_R} C_R$ for a $Z'$ boson, assuming the BW resonances satisfy $\Gamma/M < 0.3$; the light-shaded rectangle shows how that physical region is extended for a color-octet vector resonance.  For purposes of benchmark comparison, the value of $\zeta N_{S_R} C_R$ for a flavor-universal coloron ($Z'_U$) model is shown by the upper (lower) horizontal dotted line}
	\label{fig:vector-dijet}
\end{figure}

\begin{figure}[h]
	\includegraphics[width = 0.5 \textwidth]{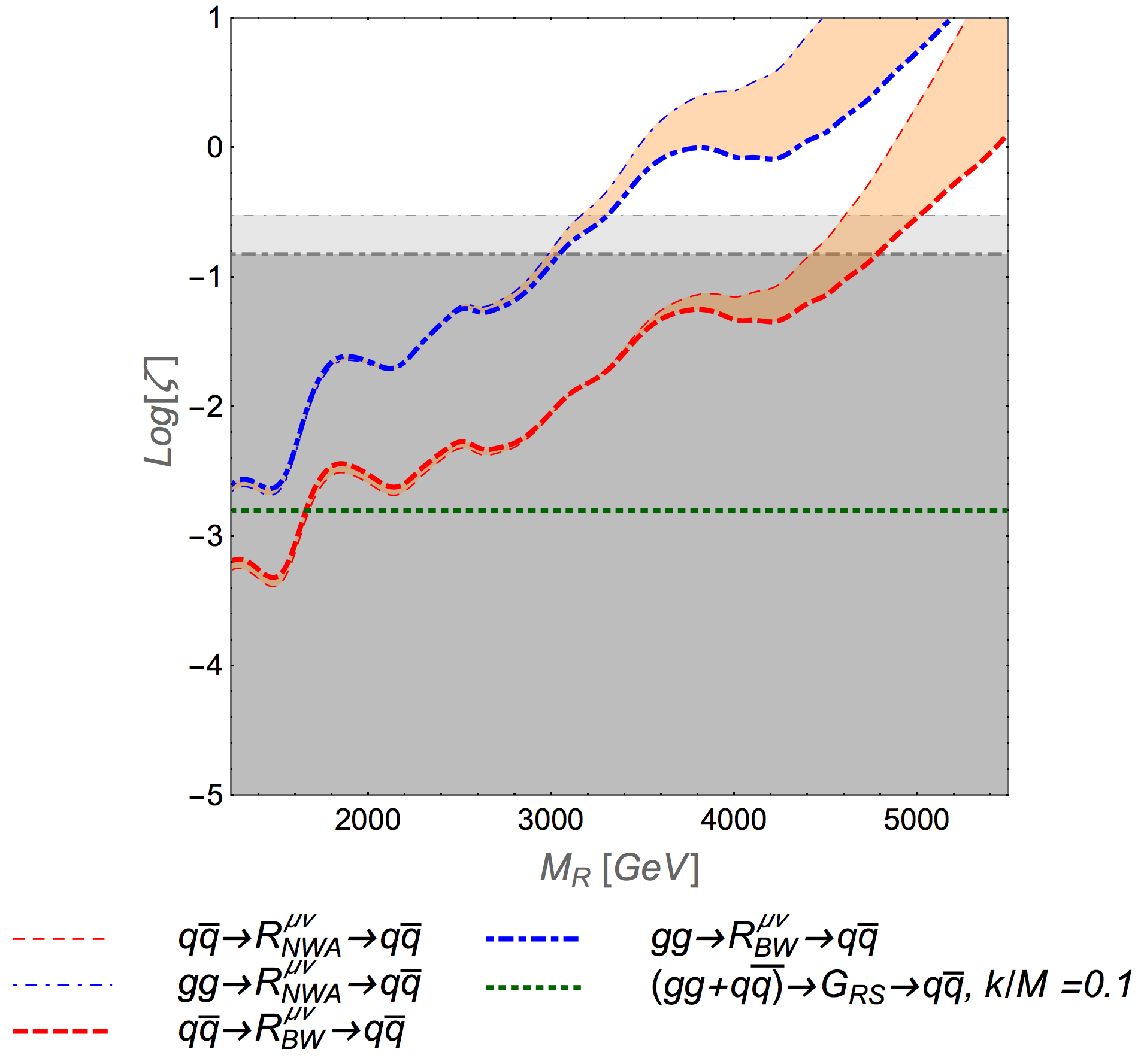}
	\caption{ \small \baselineskip=3pt Observed upper bounds (at 95\% confidence level) on spin-2 resonances from CMS dijet resonance searches ~\cite{Khachatryan:2015sja} shown in the $\zeta - M_R$ plane.  The shaded region is the area in which $\zeta$ has a physically reasonable value (see text). The upper pair of diagonal blue dashed curves is for spin-2 resonances produced via gluon fusion; the lower pair, for resonances produced via $q\bar{q}$ annihilation.  The difference between the shape of the upper pair and lower pair of curves is due to the energy dependences of the parton distribution functions for the initial-state gluons and quarks.   Within each pair of curves, the gold-shaded region indicates the difference between the upper limit in the NWA approximation (thin curve) and in the BW approximation with $\Gamma/M = 0.3$ (thick curve). The NWA yields a conservative upper limit on $\zeta$. For comparison, the horizontal green dotted line corresponds to the value of $\zeta$  for an RS graviton. } 
	\label{fig:graviton-dijet}
\end{figure}

Fig. \ref{fig:vector-dijet} shows the corresponding limits for flavor-universal\footnote{Coupling equally to $u$- and $d$-quarks, as for a resonance coupling to baryon number.} vector resonances that decay to dijets; to leading order, these are always produced via $q\bar{q}$ annihilation rather than $gg$ fusion  \cite{Zerwekh:2001uq}. As noted in the figure, we have used $N_{S_R} C_R \cdot \zeta$ as the vertical axis variable, rather than $\zeta$; 
as can be seen from Eqs. \ref{eq:gen-bran-rat} and \ref{eq:nr}, this allows us to display the limits for both color-singlet and color-octet vector-bosons via the same curves.  The dark-shaded rectangle indicates the physical region of $N_{S_R} C_R \cdot \zeta$ for a flavor-universal $Z'_U$ boson (with $\Gamma_R / M_R \leq 0.3$, $N_{S_R} = 3$, and $C_R = 1$) while the light-shaded rectangle shows how the physical region is extended in the case of a coloron (with $N_{S_R} = 3$, and $C_R = 8$). The red-shaded region between the diagonal curves illustrates the difference between the narrow width and Breit-Wigner approximations.  At low resonance masses, the NWA yields an upper limit on $N_{S_R} C_R \cdot \zeta$ that is virtually identical to the BW curve; at higher masses, the NWA computation gives a conservative, but reasonable, approximation to the BW result.  In fact, for the $Z'$ resonance, the mass range for which the BW curve diverges most strongly from the NWA curve lies outside the physical region of $ N_{S_R} C_R \cdot \zeta$.

Finally, Fig. \ref{fig:graviton-dijet} illustrates the upper limits for spin-2 resonances produced either through gluon fusion (upper curves) or quark annhilation (lower curve), and the gold-shaded region illustrates the differences between the NWA and BW calculations. The dark-shaded rectangle shows the physical region of $\zeta$ for a spin-2 resonance produced via $gg$ fusion and decaying to $q\bar{q}$, while the light-shaded rectangle shows how the physical region is extended when the intial and final states are both $q\bar{q}$. The limits on spin-2 states produced via both these channels would lie between these extremes.

Once again, the NWA yields limits that are more conservative than those assuming a Breit-Wigner form for the resonance.  At lower resonance masses, where the experimental constraints fall within the physically reasonable range of $\zeta$, the NWA and BW results are virtually identical.  At higher resonance masses, where the BW constraints start to become significantly stronger than those from the NWA, both sets of constraints eventually become too weak to limit the physical region of $\zeta$.  This plot also illustrates the same pattern noted earlier, whereby the BW curve drops visibly below the NWA curve at a lower resonance mass for states produced via $gg$ fusion compared with those produced via $q\bar{q}$ annihilation; again, this is due to the behavior of the parton luminosity functions.

It is informative to compare the results for resonances with differing spins that are produced through the same initial state partons, and therefore incorporate the same parton luminosity functions. For instance, the exclusion curves in Fig.~\ref{fig:vector-dijet} and the lower pair of exclusion curves in Fig.~\ref{fig:graviton-dijet} both show results for the $q\bar{q} \to R \to q\bar{q}$ channel; a spin-1 (spin-2) resonance is studied in Fig.~\ref{fig:vector-dijet} (\ref{fig:graviton-dijet}).  As noted in Appendix C, the acceptances for the two different spin states are quite similar in this channel.  So we expect that the exclusion curves should have the same shape (due to the same ${\cal L}_{ij}$) and be vertically displaced from one another.  More precisely, since the vertical axis variable for Fig.~\ref{fig:vector-dijet} is $N_{S_R} C_R\cdot \zeta$ while that for Fig.~\ref{fig:graviton-dijet} is $\zeta$, and since $N_{S_R} C_R = 5$ for the spin-2 resonance, we would expect the $q\bar{q}$  curve in Fig.~\ref{fig:graviton-dijet} to lie $\log 5 \approx 0.7$ below its analog in Fig.~\ref{fig:vector-dijet}. Indeed, this is what we observe.

%%%%%%%%%%%%%%%%%%%%%%%%%%%%%%%%
%  DISCUSSION
%%%%%%%%%%%%%%%%%%%%%%%%%%%%%%%%

\section{Conclusions}

A ``simplified limits" analysis of hadron collider data \cite{Chivukula:2016hvp} casts resonance search results in terms of the variable $\zeta$, defined in Eq.~\ref{eq:gen-bran-rat}, by exploiting the fact that the new physics cross-sections actually depend (to a good approximation) only on the production and decay (signal) modes considered. Using this framework, one can easily understand whether any resonance with a particular dominant production and decay channel could possibly produce a signal at the LHC matching any observed excess. Once a viable class of models has been identified, the degree to which any given theory within that class matches the observed excess can then be easily found, as it depends only on the width and branching-ratios of the resonance. 

The original simplified limits framework employed the narrow width approximation. In this paper, we examined how allowing for a finite width of the resonance modifies the simplified limit bounds extracted from LHC searches. We did so by comparing the simplified limit bounds obtained in the narrow width approximation and at finite width with the resonance described using the Breit-Wigner approximation. In particular, we illustrated applications to data from recent LHC searches covering a variety of different incoming partons, resonance properties, and decay signatures:
\begin{itemize}
\item dilepton resonances \cite{ATLAS-CONF-2016-045}, which yield the limits illustrated in Figs. \ref{fig:simlim-zprime} and \ref{fig:simlim-zprime-uudd},
\item diboson ($WZ$) resonances \cite{ATLAS-CONF-2016-055}, with the bosons decaying to dijets, deriving the results shown in Fig. \ref{fig:simlim-Wprime},
\item and dijet resonances \cite{Khachatryan:2015sja}, whose implications for particles of various spins and colors are shown in Figs. \ref{fig:scalar-dijet}, \ref{fig:vector-dijet}, and \ref{fig:graviton-dijet}.
\end{itemize}

We have demonstrated that it is straightforward to extend the simplified limits methodology to resonances with finite width.  Moreover, we found that the simplified limits derived in the narrow width approximation yield reasonable, and usually somewhat conservative (less stringent) bounds on the model parameters, compared to limits obtained by incorporating the resonance's finite width. We have enumerated limitations and possible extensions of our approach. We have shown that the simplified limits framework remains extremely valuable in the early stages of evaluating and interpreting new collider data -- and is not restricted to the case of narrow resonances.

\section*{Acknowledgments}

 The work of. R.S.C., K.M., and  E.H.S. was supported by the National Science Foundation under Grant PHY-1519045.  
 R.S.C. and E.H.S. also acknowledge the hospitality of the Aspen Center for Physics, which is supported by National Science Foundation grant PHY-1607611, during work on this paper. 
P.I. is supported by Research Grant for New Scholar Ratchadaphiseksomphot Endowment Fund, Chulalongkorn University.

%%%%%%%%%%%%%%%%%%%%%%%%%%%%%%%%
%  APPENDIX
%%%%%%%%%%%%%%%%%%%%%%%%%%%%%%%%

\section*{Appendix}

\subsection{Dilepton Selection Criteria}

Here, we summarize the experimental event selection criteria used in the ATLAS analysis~\cite{ATLAS-CONF-2016-045} of dilepton final states at $\sqrt{s}=13 ~\text{TeV}$.  This applies to our study of spin-1 resonances decaying to dileptons in section \ref{app:dileptons}.
\begin{itemize}
	\item For electrons, the pseudo-rapidity satisfies $|\eta| < 2.47$, with the transition region between central and forward regions excluded ($1.37\le |\eta| \le 1.52$). For muons the pseudo-rapidity $|\eta|<2.5$ and the region $1.01\le |\eta|\le 1.10$ is excluded.
	\item Electron discriminant variable ($95-96 \%$ efficiency) as well electron isolation requirements($99 \%$ efficiency) are used.
	\item Muon isolation requirements.
	\item Electron $E_T > 17 \text{GeV}$. Muon $p_T$ thresholds of $26~\text{GeV}$ and $50~\text{GeV}$ are used.
	\item Efficiency of triggers for a sample of $Z^{\prime}_{\chi}$ ($M_{Z^{\prime}_{\chi}} = 3~\text{TeV}$): $87\%$ for dielectron and $94\%$ for dimuon channel. 
	\item Further $E_T(p_T) 30~\text{GeV}$ for electron (muon) pair. Data derived corrections  + smearing.
	\item  Representative values of the total acceptance times efficiency for ($M_{Z^{\prime}_{\chi}} = 3~\text{TeV}$) 
	are $73\%$ in the dielectron channel and $44\%$ in the dimuon channel.
%	\item Highest invariant mass observed at $2.38$~TeV (dielectron) and $1.98$ TeV (dimuon).
\end{itemize}

\subsection{WZ Selection Criteria}

Here, we summarize the experimental event selection criteria used in the ATLAS analysis~\cite{ATLAS-CONF-2016-055} with $ 15.5\invfb$ at $\sqrt{s} = 13$~TeV.  This applies to our study of $W'$ resonances decaying to dibosons in section \ref{app:dibosons}.
\begin{itemize}
	\item Large $R=1.0$ jets are identified and after a trimming and subjet identification procedure, the trimmed jets are required to have $p_{T,J}> 200$~GeV, $m_{J}>30$~GeV and $|\eta|<2.0$.
	\item Boson ($W$ or $Z$) jets are identified using a boson tagging procedure that uses two selection criteria, namely $m_J$ and a variable $D_2^{(\beta =1)}$ that can be used to measure the compatibility of a two-prong decay topology.
	$|m_{W/Z} - m_J|< 15$~ GeV the second criterion requires a $p_T$ dependent selection of $D_2^{(\beta =1)}$.
	The boson-tagging algorithm is configured so
	that the average identification efficiency for longitudinally polarised, hadronically decaying W or Z bosons
	is $50\%$. This tagging selection reduces the multi-jet background by a factor of approximately 60 per jet.
	\item Further discrimination between boson and background jets is achieved by requiring that $N_{trk} < 30$, where
	$N_{trk}$ is defined as the number of charged-particle tracks pointing to the primary vertex3 with $p_T > 0.5$~GeV.
	\item leptonic decay modes of W and Z are rejected.
	\item Events are required have two trimmed jets with $p_{T,J}> 450$~GeV for the leading jet (to ensure full trigger efficiency) and $m_{JJ}> 1 TeV$ to avoid distortions to the mass spectrum from the $p_{T,J}$ cut.
	\item Small rapidity separation for jets $\Delta y _{12}< 1.2$ for the leading jets (to reduce t-channel backgrounds).
	\item $p_T$ asymmetry $\mathcal{A} = \frac{p_{T,J1} - p_{T,J2}}{p_{T,J1} + p_{T,J2}< 0.15}$. The signal efficiency for this
	requirement is very high, e.g. the efficiency for a HVT $W^{\prime}$ signal with a mass of $2.1$~TeV is approximately	$97\%$.
\end{itemize}

\subsection{Dijet Selection Criteria}

Here, we summarize the experimental event selection criteria used in the CMS results on dijets with $20\invfb$ of $8$~TeV data~\cite{Khachatryan:2015sja}. This applies to our study of various resonances decaying to dijets in section \ref{app:dijets}.
\begin{itemize}
	\item $p_{T_{j}}> 30$~GeV.
	\item $2\eta^*=\eta_1 -\eta_2 > 1.3$.
	\item $|\eta_{j}| < 2.5$
	\item $m_{jj} > 890$~GeV.
\end{itemize}

In the narrow width approximation and for the range of interest of resonance masses $(1250<M_R<5500)$~ GeV, it is straightforward to determine the acceptance of these cuts by simply integrating the appropriate normalized Wigner-d functions. We obtain the following values of acceptance $(A^{spin})$ for resonances with various spins:
\begin{align}
&A^0\simeq 0.57,\quad A^1\simeq 0.47,\quad A^{1/2} \simeq 0.57 &\nonumber \\
&A^2(q\bar{q} \to q\bar{q}) \simeq 0.54,\quad A^2(gg \to q\bar{q}) \simeq 0.69,\quad A^2(gg \to gg) \simeq 0.3.&
\end{align}
For broader resonances, signal events are sometimes subject to the additional requirement $m_{jj}< 1250$ .  This can cause small deviations in the values of the acceptances; since they are small, we have neglected them in our analysis.

\bibliographystyle{utphys}
\bibliography{broad_simlim_refs}

\end{document}